**Attosecond Control of Relativistic Electron Bunches using Two-Colour Fields**

M. Yeung[1,*], S. Rykovanov[2], J. Bierbach[2,3], L. Li[1], E. Eckner[3], S. Kuschel[2,3], A. Woldegeorgis[2,3], C. Rödel[2,3,4], A. Sävert[3], G. G. Paulus[2,3], M. Coughlan[1], B. Dromey[1] and M. Zepf[1,2,3]

[1]*Department of Physics and Astronomy, Queen's University Belfast, Belfast, BT7 1NN, UK*

[2]*Helmholtz Institute Jena, Fröbelstieg 3, 07743 Jena, Germany*

[3]*Institut für Optik und Quantenelektronik, Friedrich-Schiller-Universität Jena, Max-Wien-Platz 1, 07743 Jena, Germany*

[4]*SLAC National Accelerator Laboratory, 2575 Sand Hill Road, Menlo Park, California 94025, USA*

*email: m.yeung@qub.ac.uk

**Energy coupling during relativistically intense laser-matter interactions is encoded in the attosecond motion of strongly driven electrons at the pre-formed plasma-vacuum boundary. Studying and controlling this motion can reveal details about the microscopic processes that govern a vast array of light-matter interaction physics and applications. These include research areas right at the forefront of extreme laser-plasma science such as laser-driven ion acceleration[1], bright attosecond pulse generation[2,3] and efficient energy coupling for the generation and study of warm dense matter[4]. Here we demonstrate attosecond control over the trajectories of relativistic electron bunches formed during such interactions by studying the emission of extreme ultraviolet (XUV) harmonic radiation. We describe how the precise addition of a second laser beam operating at the second harmonic of the driving laser pulse can significantly transform the interaction by modifying the accelerating potential provided by the fundamental frequency to drive strong coherent emission. Numerical particle-in-cell code simulations and experimental observations demonstrate that this modification is extremely sensitive to the relative phase of the two beams and can lead to significant enhancements in the resulting harmonic yield. This work also reveals that the ability to control these extreme interactions**

**with attosecond precision is an essential requirement for generation of ultra-bright, high temporal contrast attosecond radiation for atomic and molecular pump-probe experiments[5,6].**

During the rising edge of a sufficiently intense laser pulse (peak intensity $>10^{18}$ Wcm$^{-2}$), which is incident on a solid density target, a fully ionised plasma is formed such that the peak of the pulse interacts with a dense plasma surface. Electrons formed at such a surface can be driven to relativistic velocities within a single optical cycle and subsequently emit XUV radiation. For multi-cycle laser pulses, these XUV bursts are generated periodically at the driving laser frequency, such that the radiation is emitted in the form of a train of attosecond pulses which are observed in the spectral domain as high order harmonics. This emission can be modelled as a periodic Doppler upshift as the electrons move towards the incident laser at velocities close to the speed of light, a case called the Relativistically Oscillating Mirror (ROM) model[7-9]. However, under certain conditions, the electrons form extremely short, dense nanobunches which can undergo synchrotron-like trajectories so that the XUV generation mechanism is more accurately modelled as Coherent Synchrotron Emission (CSE)[2].

Recent numerical simulations have suggested that using a two-colour field, consisting of the fundamental pulse and its second harmonic with the right phase relationship, can optimise the trajectories of the emitting surface electrons leading to a significant increase of the attosecond XUV yield[10]. Previous simulations[11] also showed a significant effect but focused on extremely high incident intensities and the influence of the relative phase between each driving frequency was not determined. Previous experimental studies[12-14] observed no clear enhancement, however, the fundamental beam was used as only a weak probe in a non-collinear setup and the relative phase between the two frequencies was not controlled.

Here we examine the detailed mechanism for the anticipated XUV enhancement and demonstrate, using numerical simulations and experiments, that, ultimately, attosecond control of the relative phase of the two driving fields is required to match the dense electron bunch formation with the accelerating potential of the two-colour field.

Figure 1 compares the effect of the phase mismatch between the fundamental and second harmonic on the bunch formation and harmonic emission directly. In Figures 1a-c the relative phase, $\Delta\phi$, between the fundamental field and its second harmonic is 0, while $\Delta\phi=\pi$ for Figures 1d-f. Figures 1a and d show the incident electric field profiles of both the pure and dressed driving fields (red dash-dotted and blue line respectively). Figures 1b and e show the resulting electron density and reflected electric fields for harmonic orders >20 from numerical simulations for the conditions described in Methods, while Figures 1c and f show the corresponding particle velocities.

As can be seen the emission is clearly optimised for the $\Delta\phi = 0$ case. Here, each cycle's emission originates from one primary bunch that is compressed and rapidly accelerated away from the surface to relativistic velocities. This bunch subsequently emits an attosecond pulse before breaking up and returning to the plasma. Some secondary bunches are also generated but do not contribute significantly to the emission for harmonic orders >20. This behaviour is a direct result of the formation of a steep gradient in the dressed electric field shape (labelled X in Figure 1a). In addition, the bunch is seen to emit at the moment when the particle velocities are highest. This maximises the cut-off frequency ($\sim\gamma^3$ where $\gamma$ is the Lorentz factor[9]) for the generated radiation, which demonstrates that the precisely timed two-colour field matches the XUV emission to the acceleration provided by the fundamental driving laser field.

For the worst phase ($\Delta\phi=\pi$) the acceleration stage for the electrons is staggered, leading to a lower bunch density when its velocity reaches its maximum. This is due to the formation of a saddle point in the rising edge of the driving field (labelled Y in figure 1d). Since the emitted power scales with $n_b^2$ for a bunch with electron density $n_b$, this results in weaker XUV radiation in addition to a broad temporal distribution of the emission (figure 1e). Furthermore, the electrons do not reach similarly high velocities as in the optimal case, nor is the point of maximum velocity synchronised with the emission times. These simulations illustrate the degree to which control of the relative phase of the two colours can strongly influence the sub-cycle dynamics of the electron bunches.

Next we investigate if it is possible to achieve this degree of attosecond control in the laboratory. A two-colour field was generated in a collinear setup and consisted of a 35fs, 800nm pulse with a peak normalised vector potential $a_0 \approx 1.25$ (where $a_0^2 = I\lambda^2/1.38 \times 10^{18} Wcm^{-2}\mu m^2$ for an intensity I and wavelength $\lambda$) and its second harmonic with one third of the intensity. The experimental setup is shown in figure 2 with further details given in the methods. The critical aspect of this setup is the fine control of the phase between the colours using a rotatable fused silica wafer, which allowed the relative timing to be tuned over a full cycle of the 2$^{nd}$ harmonic. The optical setup is similar to that used in ref 15.

Figures 3(a) and (b) show the measured spectral energy density, as the second harmonic phase is varied, for photon energies from 24eV to 33eV and 33eV to 42eV respectively. The higher spectral range was chosen for detailed analysis, as it is free of a competing harmonic generation mechanism known as Coherent Wake Emission[16] (CWE) which can become dominant for lower photon energies. The details of this mechanism and its relation to the current results are discussed in the supplementary information. From the data, a strong phase dependence is observed and control over the harmonic yield is demonstrated, as predicted by simulations. The fast decay of the spectra for higher orders indicates that we are close to the cut-off region[2]. This is apparent in the rescaled false colour experimental spectra of figure 3(c) where the highest order harmonics have a more critical dependence on having the correct phase. As previously discussed, this is indicative of the greater peak velocities reached in the optimum case, which is the key factor in determining the cut-off harmonic.

For the higher harmonic orders the detected signal ranges from near complete suppression to an energy of 125±20nJ across the spectral range in the strongest case. For the full spectrum from 24eV to 42eV (16th to 27th harmonic of the fundamental), an energy of 11.8±0.7μJ is observed at the optimum phase which is greater than even the best optimised HHG sources from the more widespread 3-step process in gaseous media[17] using two-colour fields, loose focusing and similar initial laser parameters[18].

It is well known that the efficiency of relativistic surface harmonic generation is strongly dependent on the scale length of the plasma density gradient[19,20]. Here, the length of the plasma density gradient was controlled by the use of a prepulse mirror (see Methods) as in the work of Kahaly *et al.*[20]. With only the fundamental beam, the density gradient was optimised using a pulse arriving 260±20fs before the main pulse. Indeed, the prepulse was essential to achieve any measureable harmonic signal at the higher photon energy range using only the fundamental laser frequency. In contrast, when the second harmonic is added, the harmonic signal was found to be optimal without any prepulse (thus yielding a much shorter scale length). Comparing these two cases, an enhancement factor of 57±10 was found for the higher spectral range when using the two-colour field demonstrating the huge gains that can be made by converting only a small fraction of the fundamental laser energy to the second harmonic. While trajectory control plays an important role in two-colour experiments[21] aiming to exploit the 3-step HHG process in gaseous media, the underlying physical mechanism behind this effect is very different for these two cases, with atomic recollisions, rather than relativistic electron bunch control, at the heart of the process. Of particular importance is the polarisation of the second harmonic which can be either parallel or orthogonal for gaseous media interactions[21-23] but must be parallel for solid targets. This was verified in the current work by rotation of the achromatic waveplate yielding no enhancement for orthogonal polarisations.

As this is a collinear setup, the shorter wavelength of the second harmonic means it is focussed to a smaller spot than the fundamental (full width at half maximum (FWHM) of 2.7μm for 2ω vs 4.7μm for 1ω). We estimate the conversion efficiency relative to the energy of both beams concentrated within the 2ω spot to be ≈2X10$^{-3}$ for the full photon energy range from 24 to 42eV (the signal present with the 1ω beam only was subtracted for this estimate). This remarkable efficiency from only a weakly relativistic driver ($a_0$≈1) highlights the potential of this scheme to bring efficient relativistic harmonic generation within the reach of a much wider range of laser systems. Applying this technique to much higher power laser systems is, therefore, an exciting prospect. Even assuming only constant conversion in the demonstrated range, using the same intensity with PW class lasers would result in XUV spectra containing mJ of harmonic energy. Scaling the intensity higher by factors of 100 and beyond is possible with PW class lasers and is expected to result in very substantial increases in yield at higher photon energies, opening up the possibility of generating extremely high energy attosecond pulses. These possibilities require experimental investigation, due to the challenge of performing fully predictive simulations of dense relativistic plasma interactions.

The enhancement in the integrated harmonic signal for orders 22 to 27 is plotted in figure 4 for both the experimental and simulation data. Here we see remarkable agreement for a 1D code, further supporting that the origin of this effect lies in the temporal variation of the incident field. Additionally, the signal obtained when there is no 2$^{nd}$ harmonic present is also plotted here as a baseline and is seen to be only very slightly better than the worst case of the two-colour pulse. The plotted phase dependence clearly demonstrates that accurate sub-cycle control over the relative timing is critical for optimising the harmonic yield.

Overall, this work demonstrates sub-laser cycle control of relativistic electron bunch trajectories by tuning of the relative phase of an added second harmonic pulse. The huge enhancements in the emitted XUV yield for modest laser peak powers (<2TW) brings this relativistic high harmonic source within the reach of an extremely wide range of lasers which can reach these powers at 10Hz repetition rates, and is not far from that achievable with current kHz systems[24]. Extending this method to even higher power systems (using, for example, annularly split beams) has the potential to produce attosecond pulses with unprecedented pulse energies.

**METHODS**

**Two-colour field setup**

The laser used for this study was the JETI 40 Ti:Sapphire system apertured to a 23mmX23mm beam to match the calcite crystal size.  This beam was contrast enhanced using a plasma mirror system[25] in order to ensure a very steep density profile that could be controlled via a translatable prepulse mirror[20].  Without conversion to its second harmonic, this setup yielded pulses of 60mJ with 35fs pulse duration and an intensity of ≈$3.6 \times 10^{18}$Wcm$^{-2}$ on target after focusing by a 15cm focal length silver coated parabola.  A potassium dihydrogen phosphate (KDP) crystal is used to frequency double the pulse with ≈10% efficiency yielding an intensity of ≈$1.1 \times 10^{18}$Wcm$^{-2}$ from the second harmonic and a small drop in intensity for the fundamental due to the energy lost from the conversion. The time delay between the two colours is precompensated by a calcite crystal before the beam passes through an achromatic waveplate which introduces a retardation of 1.5$\lambda$ for 400nm and 1$\lambda$ for 800nm in order to ensure both pulses are P-polarised on target.  Finally, the relative phase is tuned by controlling the rotation of a 500$\mu$m fused silica wafer thus changing the relative path difference.  Temporal overlap was confirmed by optimising third harmonic generation in a BBO crystal.

**Scale Length Control**

The plasma surface density gradient was controlled using the same method as ref 20, by the use of a 3mm thick fused silica substrate with an anti-reflection coating on the front side and a high reflectivity coating on the reverse. The optic is placed directly in front of one of the full beam mirrors and reflects a small section of the beam which is then focused by the same focusing optic as the main pulse.  This lower intensity spot acts as a prepulse, ionising the surface of the target making an expanding plasma. Thus, by controlling the timing of the arrival of the prepulse by translation of this special optic, the length of the expanding plasma at the time the driving pulse arrives can be controlled. The high reflectivity coating is placed on the reverse side so that the prepulse picks up enough group delay in the substrate to allow the prepulse to be initially positioned late to ensure that zero timing can be reached as is scanned earlier.

**Spectrometer**

The reflected radiation from the interaction is first filtered by a 200nm aluminium foil before it is focused by a Ni coated toroidal mirror which makes a one-to-one image of the source. This image is spectrally dispersed by a 1000 lines/mm freestanding gold transmission grating and the resulting spectrum is detected on an ANDOR Newton DO940P-BN back illuminated CCD. The spectrometer has an 8mradX6mrad collection angle which, assuming diffraction limited harmonic emission, would be expected to collect almost the entire harmonic beam. This is only valid for the relativistic harmonics if negligible denting of the target[26-28] is assumed. Without this assumption, the measured values constitute a lower limit for the actual generation efficiency. This spectrometer was previously calibrated at a synchrotron source allowing absolute energy values to be determined[29].

**PIC Simulations**

1D PIC simulations were performed using the code PICWIG[30]. For the simulations of the phase dependence of the harmonic yield, the cell width used was $\lambda/500$ ($\lambda$=laser wavelength) and the target had a $\lambda/40$ exponential density ramp at the front of the target leading up to a peak density of $400n_c$ where $n_c=\omega^2 m_e \varepsilon_0/e^2$ is the critical density for an electromagnetic wave with angular frequency $\omega$ and with $m_e$, $\varepsilon_0$ and e equal to the electron mass, electric permittivity and electron charge respectively. Ions were mobile and each target cell contained 1000 electron and 1000 ion macroparticles. The laser pulse was modelled as a $\sin^2$ pulse with a total width of 36.8 laser cycles and peak $a_0$ values of 1.25 for the fundamental and 0.36 for the 2$^{nd}$ harmonic. For the simulations yielding the electron density and trajectory data, a shorter pulse (10 cycles) and a cell width of $\lambda/1000$ was used for improved spatial resolution and the number of particles per cell was reduced to 500 for each species to manage the higher computational load.

**ACKNOWLEDGEMENTS**

This work was supported by the European Regional Development Fund (EFRE) and by Deutsche Forschungsgemeinschaft (SFB TR18-A7). B.D. acknowledges support from the EPSRC through grants and a Career Acceleration Fellowship. S.R. acknowledges the support by the Helmholtz Association (Young Investigator's Group VH-NG-1037). C.R. acknowledges support from the VolkswagenStiftung. The authors acknowledge the support of Falk Ronneberger and Burgard Beleites as laser operators.

**AUTHOR CONTRIBUTIONS**

All authors contributed to the planning and implementation of the experiment. M.Y., S.R. and M.Z. carried out the data analysis and the PIC simulations were performed by S.R.

**COMPETING FINANCIAL INTERESTS**

The authors declare no competing financial interests

**Figure 1**

**Two-colour field driven attosecond pulse emission** – Comparison of simulation data for the optimum (a-c) and worst (d-f) relative phase of the 2$^{nd}$ harmonic pulse. The fundamental pulse had a peak $a_0$=1.25 whilst the 2$^{nd}$ harmonic peak $a_0$=0.36. Here, $T_L$ represents the pulse duration, $\lambda_L$ is the fundamental wavelength and c is the speed of light. The incident fields for each case are plotted and compared with that of the fundamental only in (a) and (d). The evolution of the electron density over two laser cycles is shown in (b) and (e) along with the transverse electric field filtered to remove frequencies at the 20$^{th}$ harmonics and below to illustrate the attosecond pulse emission. Colourbars are shown below. The magnified region shows the formation, acceleration and emission of the first bunch in this window. The maximum velocity reached by the electrons is indicated by the gradient of the dashed line in this magnified view. Additionally, (c) and (f) show the velocity against time for each particle and the time at which this is maximum for the first bunch is indicated, in each case, by a dotted line.

**Figure 2**

**Experimental setup** – A schematic of the optical setup for generating the two-colour field and the spectrometer layout. The timing and polarisation of each frequency is indicated after each stage. The p-polarised fundamental is frequency doubled in a KDP crystal, after which the orthogonally polarised second harmonic is brought early using a calcite time delay compensator. The two colours are made to have parallel polarisation using an achromatic waveplate before passing through a fused silica wafer which can be tilted in order to control the relative phase. After focusing to high intensity on target, the reflected radiation is collected by a toroidal mirror which, after spectral dispersion by a gold transmission grating, is imaged onto an XUV CCD.

**Figure 3**

**Phase dependence of harmonic spectra** - The measured calibrated spectra, each averaged over 10 shots, are plotted here for each relative phase setting. The spectra for harmonics 16 to 21 are shown in (a) whilst (b) shows spectra for harmonics 22 to 27. The zero phase position is determined from comparisons with simulations. A false colour map is shown in (c) where the data has also been normalised and scaled to make the higher orders visible to show the full spectral dependence on the relative phase.

**Figure 4**

**Comparison with simulation** – The enhancement, relative to the single colour driver, in the total signal integrated over harmonics 22 to 27 (blue crosses) is plotted against the relative phase alongside results from 1D PIC simulations (red circles). The phases are absolute and taken from the simulation parameters while the phase axis for the experimental data has been set to match the simulation results. Additionally, the simulation maximum has been scaled to match the experimental data. The shaded region indicates the standard deviation of the experimental data. The lower shaded orange region shows the same variation of the signal in this spectral range for the case when there is no $2^{nd}$ harmonic present and the prepulse position has been optimised.

Figure 1

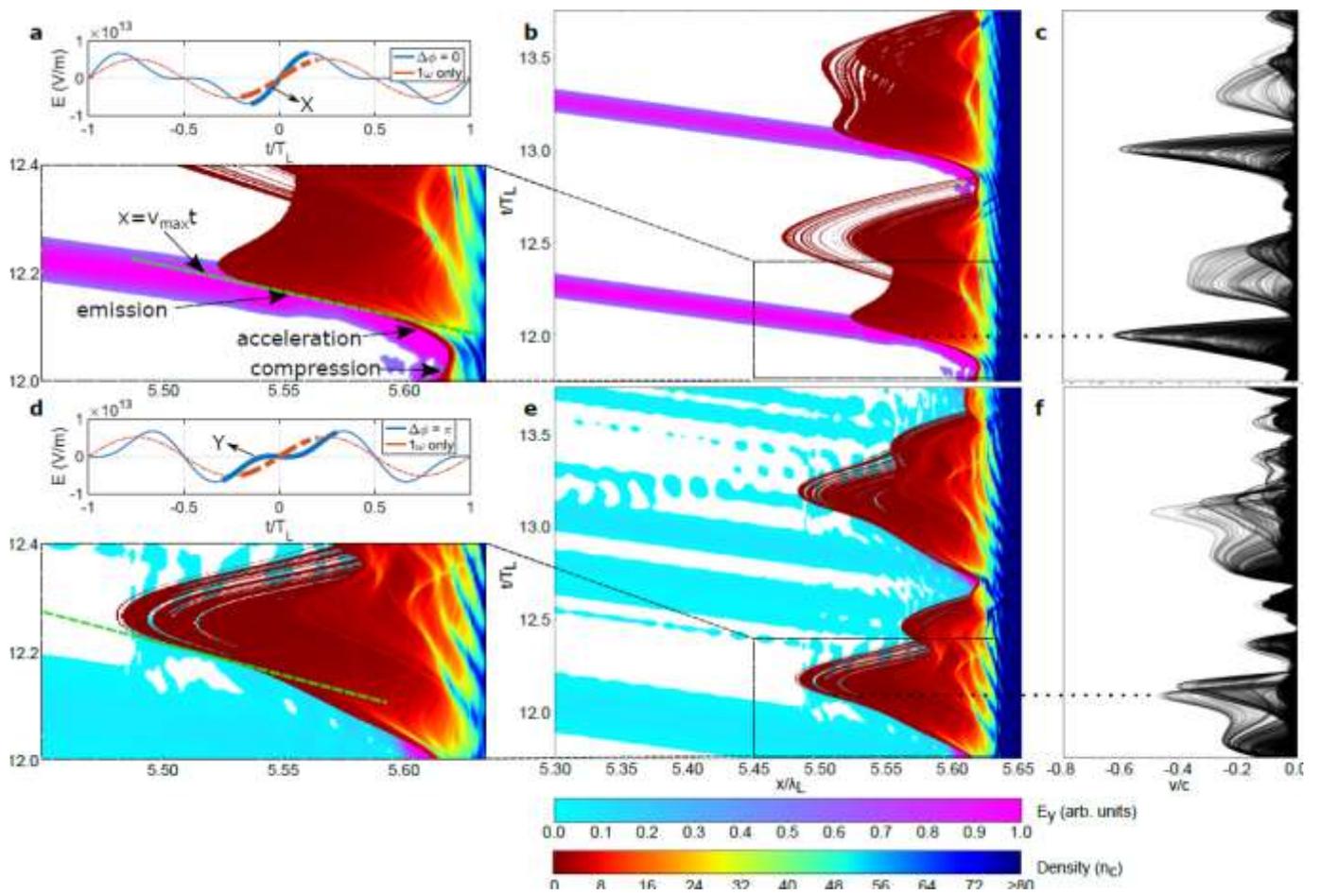

Figure 2

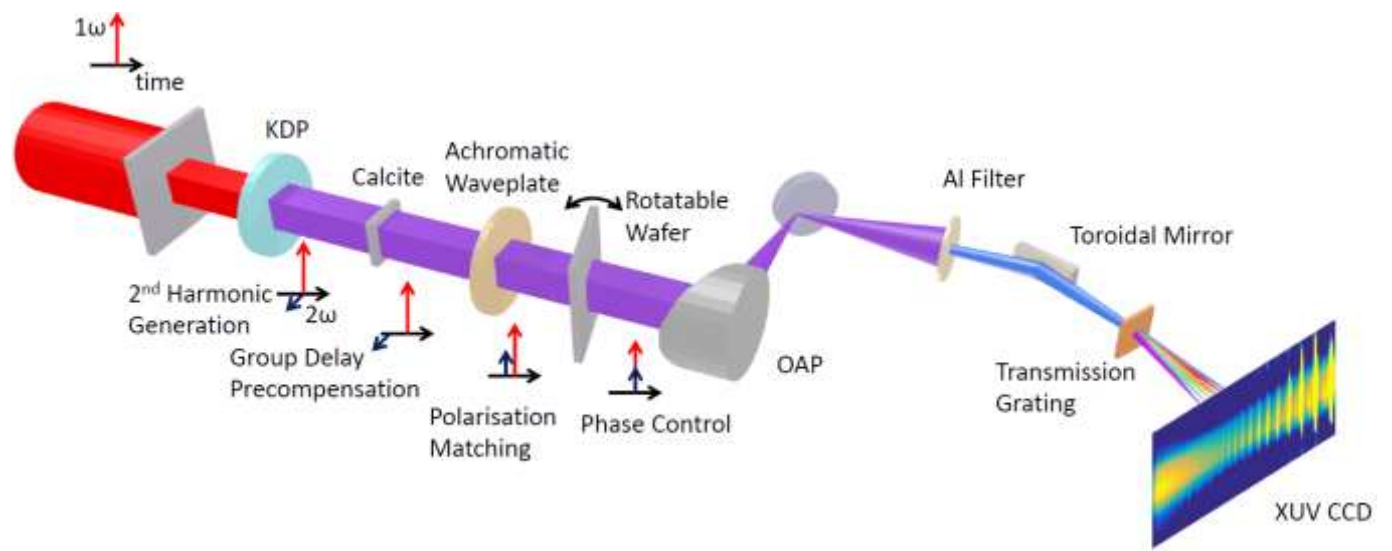

Figure 3

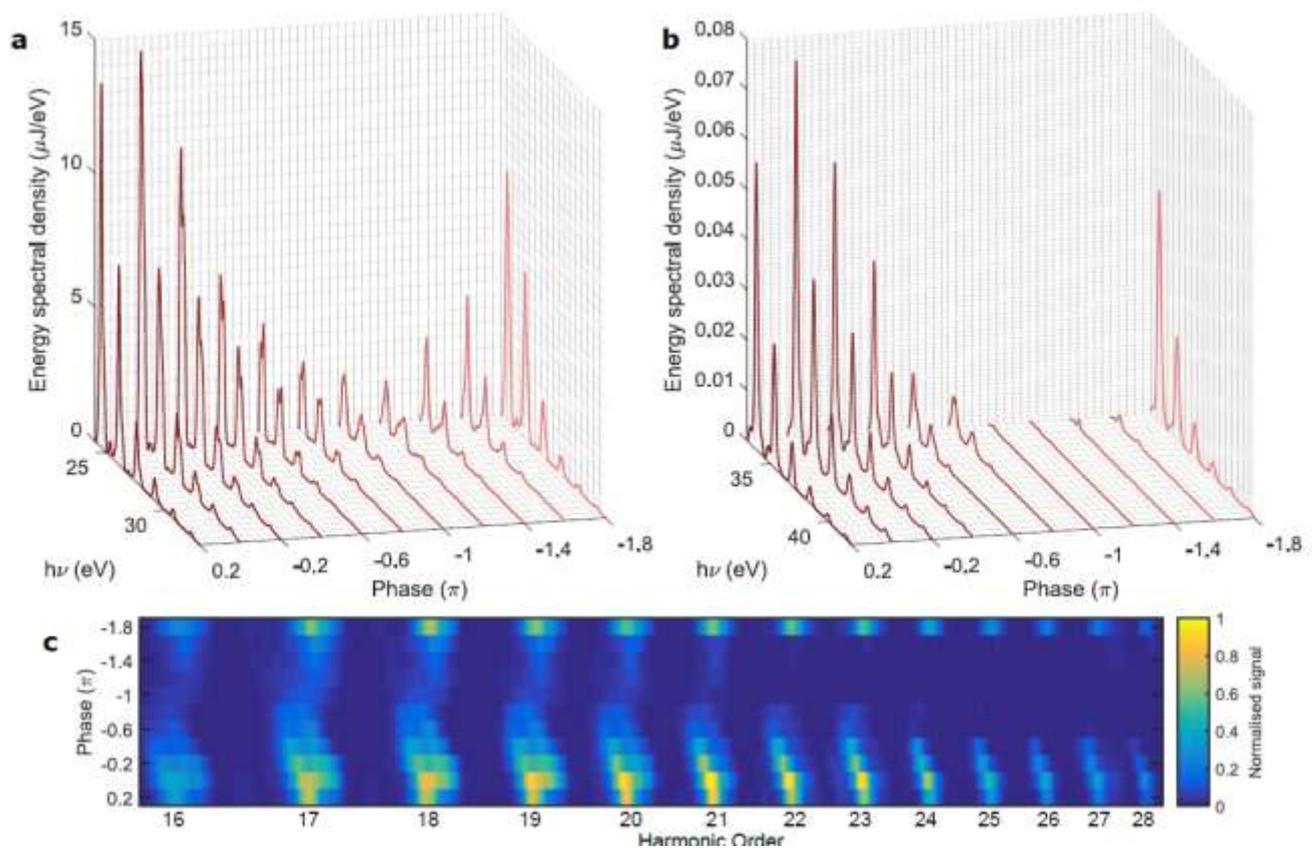

Figure 4

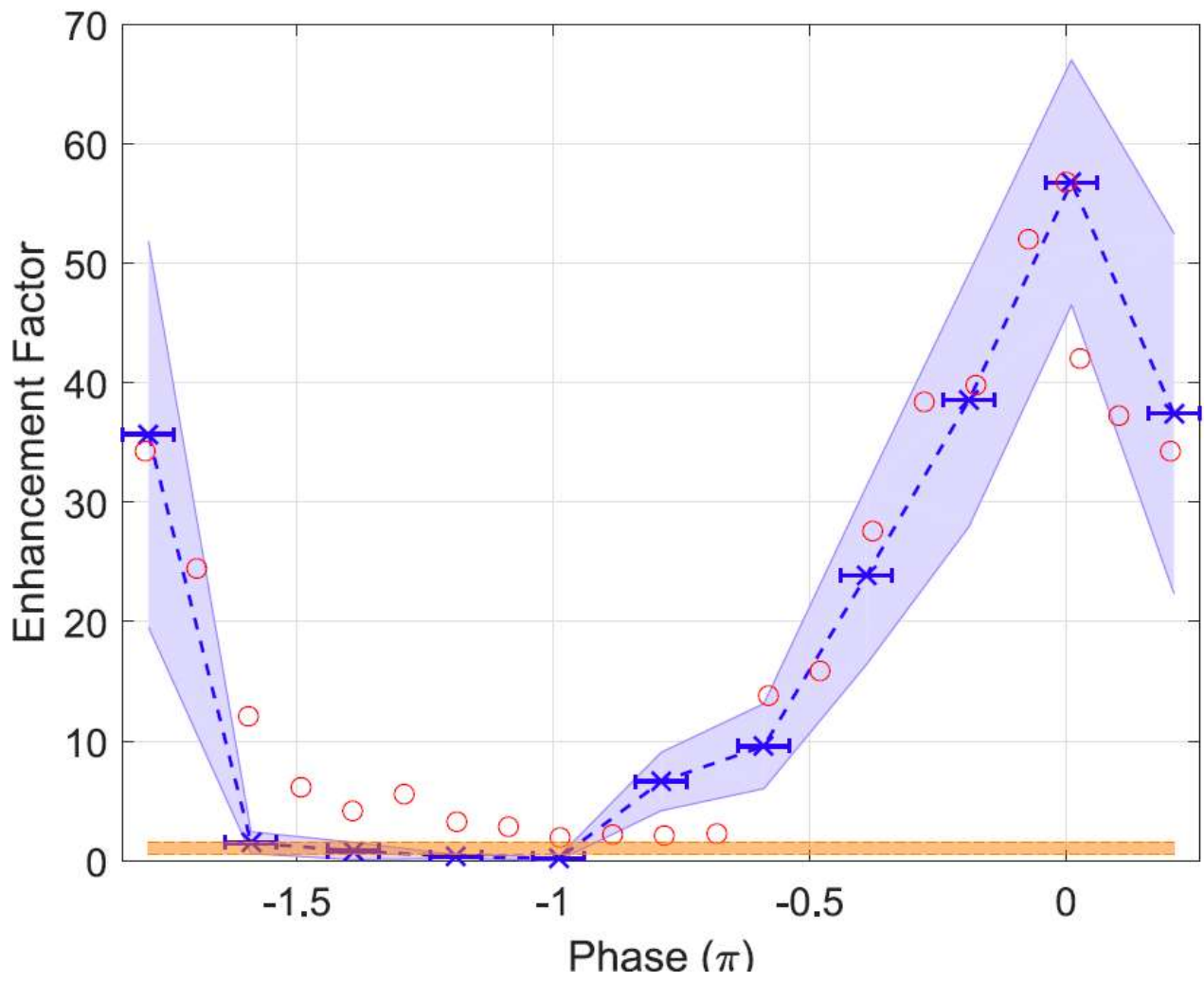